\documentclass{aa}

\usepackage{graphicx}
\usepackage{txfonts}
\usepackage[utf8]{inputenc}
\usepackage{mathtools}
\DeclarePairedDelimiter\abs{\lvert}{\rvert}
\usepackage{graphics}
\usepackage{xcolor}
\usepackage{tikz}
\usetikzlibrary{patterns.meta}

\usepackage{amsmath}
\usepackage{amssymb}
\usepackage{feynmp-auto}
\usepackage{epsfig}
\usepackage{caption}
\usepackage{subcaption}
\usepackage{hyperref}
\usepackage{latexsym}
\usepackage{pdfpages}
\usepackage{setspace}
\usepackage{threeparttable}
\usepackage{float}
\usepackage[version=4]{mhchem}
\usepackage{siunitx}
\usepackage{hyphenat}
\usepackage{booktabs}
\usepackage{comment}
\usepackage{orcidlink}

\graphicspath{ {./images/} }

\usepackage{hyperref}
\hypersetup{
    colorlinks=true,
    linkcolor=blue,
    filecolor=blue,      
    urlcolor=blue,
    citecolor=blue
    }

\definecolor{codegreen}{rgb}{0,0.6,0}
\definecolor{codegray}{rgb}{0.5,0.5,0.5}
\definecolor{codepurple}{rgb}{0.58,0,0.82}
\definecolor{backcolour}{rgb}{0.95,0.95,0.92}

\begin{document}

\title{Intermediate-mass black holes and contribution to extragalactic background light from Population III stars in \\ Milky Way-like galaxies}


\author{A. Mkrtchyan \inst{1}\thanks{Corresponding author: \href{mailto:artak.mkrtchyan@uni-hamburg.de}{artak.mkrtchyan@uni-hamburg.de}}\orcidlink{0000-0002-6217-8539} \and D. Horns \inst{2}\orcidlink{0000-0003-1945-0119}}



\institute{Universität Hamburg, Institut für Quantenphysik, Luruper Chaussee 149, 22761 Hamburg, Germany \and Universität Hamburg, Institut für Experimentalphysik, Luruper Chaussee 149, 22761 Hamburg, Germany \\}

\date{Received date /  Accepted date}

 
\abstract{The mass range of observed black holes extends from stellar-mass to supermassive scales, yet the existence of objects in the intermediate-mass range of $10^{2} - 10^{5} \SI{}{M_{\odot}}$ remains unconfirmed. 
Black holes are suspected to compress the surrounding dark matter distribution, forming a ``spike''. If dark matter is self-annihilating, the ``spike'' could produce gamma-ray emission sufficiently luminous to be detected. This work aims to estimate the number of expected unmerged intermediate-mass black holes in a Milky Way-like galaxy that could form such spikes. These intermediate-mass black holes are assumed to have formed from the collapse of high-mass Population III stars, such that the resulting merger rate is constrained by observations of gravitational wave emission. 
It is furthermore estimated to what extent the progenitor Population III stars contribute to the extragalactic background light. 
The Population III stars are simulated and tracked using the A-SLOTH semi-analytical simulation code and the resulting number of intermediate-mass black holes is constrained by applying the Population III binary black hole merger rate to an effective volume determined from the Population III star formation rate. 
In this framework, $\sim 130$ unmerged IMBHs from Population III stars are expected to reside in a Milky Way-like galaxy. The contribution  of their progenitors to the extragalactic background light in the near-infrared is less than $10^{-3}\SI{}{\nano\watt\per\square\metre\per\steradian}$, well below previous estimates. 
}


\keywords{stars: black holes -- stars: Population III -- gravitational waves -- (cosmology:) cosmic background radiation}

\titlerunning{Population III remnants in Milky Way-like halos}
\authorrunning{Artak Mkrtchyan \& Dieter Horns}

\maketitle

\section{Introduction} \label{section:introduction}

\begin{table*}[t]
    \centering
    \begin{tabular}{ccccccc} \toprule
                SFR   & $a$ & $b$ & $c$ & $d$ & $e$ & $z_{0}$ \\ \midrule
        $\psi_{2}$ & \SI{-4.4211e+00}{} & \SI{0.13755e-00}{} & \SI{-3.8859e-02}{} & - & - & \SI{-7.0523e+00}{} \\ 
        $\psi_{4}$ & \SI{-12.382e+00}{} & \SI{1.2428e+00}{} & \SI{-6.1530e-02}{} & \SI{1.5012e-03}{} & \SI{-1.7079e-05}{} & \SI{2.1074e+00}{} \\ \bottomrule
    \end{tabular}
    \caption{Parameters for A-SLOTH SFR fit functions.}
    \label{table:fit_params}
\end{table*}

The mass distribution of observed Black Holes features a population of stellar-mass black holes at the low mass end and a population of supermassive black holes (SMBHs) at the high mass end. In the framework of hierarchical structure formation, supermassive black holes are expected to form through repeated merging and accretion of matter from a seed population of lighter (stellar-mass) black holes. In this picture, objects in the intermediate mass range with $10^{2} - 10^{5}$ \SI{}{M_{\odot}}  are naturally expected to exist but remain to be  conclusively identified. 
However, there are several observations that are possibly explained by the existence of Intermediate-Mass Black Holes (IMBHs). The following main points of motivation are often discussed:

\begin{enumerate}
    \item Observations of larger-than-expected velocity dispersions in select globular clusters hint at central IMBHs ``stirring up'' the stars and introducing additional contributions to the velocity dispersion \citep{1976MNRAS.176..633F}. An analysis on the stellar kinematics of the globular cluster $\omega$ Centauri by \citet{Noyola_2008} reveals an increase of the mass-to-light ratio towards the center of the cluster and suggests the existence of a central BH with a mass of \SI{4e4}{M_{\odot}}. A study by \citet{H_berle_2024} has identified seven fast-moving stars in the central 3 arcseconds of $\omega$ Centauri, with velocities exceeding the escape velocity of the cluster, indicative of a central BH with a lower bound mass of \SI{8200}{M_{\odot}}.

    \item Another point of motivation comes from the observation of Ultra-Luminous X-ray sources (ULXs). These are sources that emit X-Rays with luminosities of $L_{X} \sim 10^{39} - 10^{40} \SI{}{erg/s}$, exceeding the luminosities of accreting stellar BHs at the Eddington limit while being below the luminosities generated by typical Active Galactic Nuclei (AGN) by a margin of several orders of magnitude. Common explanations for ULXs are given by X-ray binaries \citep{2001ApJ...552L.109K,Lasota_2023}, although accreting IMBHs provide an alternate explanation for select sources \citep{Miller_2003}.

    \item Possibly the strongest point of motivation for the existence of IMBHs comes from the rapid growth of SMBHs and the great masses they managed to acquire in the early universe. \citet{bogdan2023evidence} reported the observation of the high redshift quasar UHZ1 with an estimated mass of $\sim 10^{7} - 10^{8}$ \SI{}{M_{\odot}} using the Chandra X-ray observatory. The object was later spectroscopically confirmed to have a redshift of $z = 10.1$ using the James Webb Space Telescope (JWST) \citep{goulding2023uncover,natarajan2023detection}. With having only few $\SI{100}{Myr}$ to grow, IMBHs with $\sim 10^{4} - 10^{5} \SI{}{M_{\odot}}$ are thought to act as seed BHs in the early universe that can reach supermassive scales by mere accretion of matter.
\end{enumerate}

Irrespective of mass, BHs are expected to modify the Dark Matter (DM) distribution and cause the adiabatic growth of a DM overdensity, called a ``spike''. If DM is self-annihilating, the resulting gamma-ray luminosity of the spike will be enhanced in comparison to an unmodified DM mass density profile \citep{Zhao_2005,Bertone_2005}, increasing the reach for indirect searches with gamma-ray telescopes as demonstrated by \citet{Aharonian_2008}. Indications for the existence of these types of overdensities are drawn from rapid orbital decay rates observed in two low-mass X-ray binary systems (LMXBs) which are suggestive of the companion stars experiencing dynamical friction in a DM spike environment \citep{Chan_2023}. With IMBHs dressed with a DM spike providing another way to probe DM, there is interest in studying the IMBH population itself \citep{aschersleben2024gamma}. 

This work aims to estimate the number of IMBHs in Milky Way-like galaxies that might be dressed with DM spikes. Specifically, unmerged IMBHs are of interest, as the spike might be disturbed or disrupted during a merger with another BH. With the merging behavior of stellar Binary BHs (BBHs) studied observationally using gravitational wave detectors, they provide constraints on the BH-forming stellar population. For this reason, the presented work considers the formation of $10^{2} - 10^{3}$ M$_{\odot}$ ``light seed'' IMBHs following the collapse of high-mass Population III (Pop III) stars. In addition to being IMBH progenitors, Pop III stars are expected to be short-lived and luminous and so might have left detectable footprints in present-day stellar observables, although latest efforts have deemed this possibility unlikely (\citep{Sun_2021}). 
It is therefore tested to what extent an IMBH forming population of Pop III stars contributes to the optical and infrared bands of the diffuse cosmic background radiation called the Extragalactic Background Light (EBL) \citep{Hill_2018}.

The methods for simulating Pop III stars and for assessing the merging behavior of their remnant BBHs are discussed in Sect. ~\ref{section:methods}. The primary result on the number of unmerged IMBHs in Milky Way-like galaxies and the implications of the required Pop III stellar population on stellar observables are presented in Sect. ~\ref{section:results}. In Sect. ~\ref{section:discussion}, caveats and uncertainties of this work are discussed and compared with previous work conducted on Pop III IMBHs.

\section{Methods} \label{section:methods}

\subsection{Simulation of Population III Stellar Numbers} \label{section:simulation}
In order to estimate the number of unmerged IMBHs from Pop III stars in Milky Way-like galaxies, it is necessary to track the evolution of their progenitors. For this purpose, the A-SLOTH code has been used \citep{Hartwig_2022}. It is a highly efficient and parallelized Semi-Analytical Model (SAM) that is calibrated with six local observables of the Milky Way. Sophisticated implementations of feedback mechanisms governing star formation and treatment of metallicity allow for individual sampling and tracking of Pop III and Pop II stars until their supernova phases. In this work, A-SLOTH Version 1.2.1 was used \footnote{Within the time of this work, version 1.3.0 of the A-SLOTH code was released (\cite{Hartwig_2024}). This release contains a greater degree of calibration, utilizing more observables, and bugfixes that have only negligible effects on prescriptions of physical processes in the model. While the new best-fit model assumes a steeper Pop III initial mass function (IMF) than the previous model indicated, resulting in the formation of fewer high-mass Pop III stars, the authors note that this quantity is weakly constrained and a log-flat IMF, forming more high-mass Pop III stars cannot be excluded. This, in effect, turns the results from this work into an upper limit estimation with respect to version 1.3.0 of A-SLOTH.}. Since in its default configuration, the simulation code aims to reproduce the Milky Way, the set of simulation parameters was largely left unchanged. Following recent numerical simulations (\cite{klessen2023}), the Pop III Initial Mass Function (IMF) was chosen to be log-flat with the following functional form:

\vspace{-0.1cm}

\begin{equation}
    \xi(m) = \frac{\mathrm{d}N}{\mathrm{d}m} = N_{0} \, m^{-\gamma},
\end{equation}

\vspace{0.3cm}

\noindent where $\gamma = 1$ is the power-law index and $N_0$ is a normalization factor such that: 

\vspace{-0.1cm}

\begin{equation} \label{unity_condition}
    1 = \int\limits_{m_{min}}^{m_{max}} m \, \xi(m) \, \mathrm{d}m.
\end{equation}


\begin{figure}[h]
    \centering
    \includegraphics[scale = 0.35]{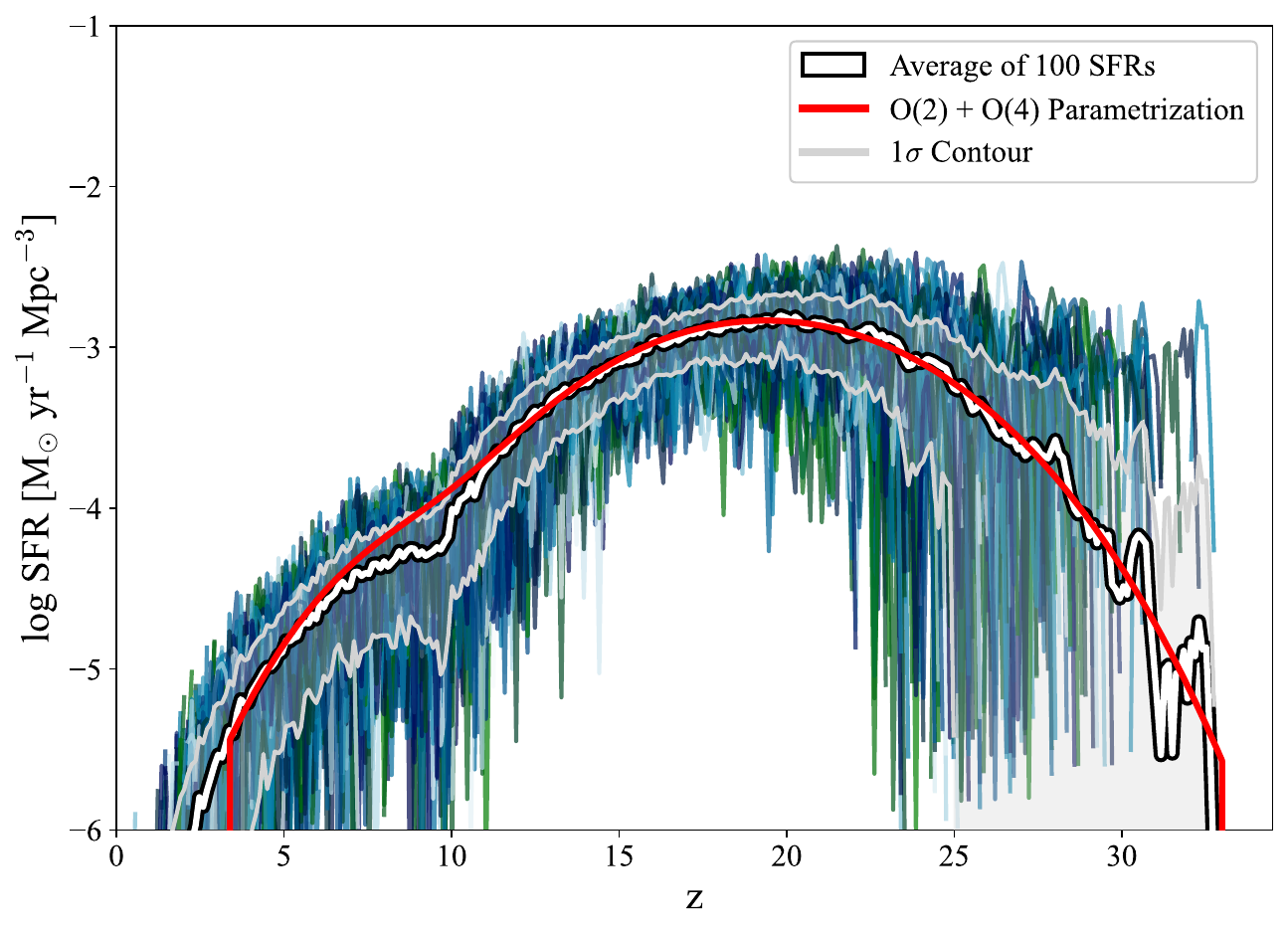}
    \caption{SFRs from 100 A-SLOTH simulations (blue and green lines). The averaged SFR is shown in white with the overlayed parametrization in red.}
    \label{figuire:ASLOTH_SFR}
\end{figure}

\noindent To allow the formation of high-mass BHs, the stellar mass range was set to $\SI{5}{M_{\odot}} - \SI{500}{M_{\odot}}$ and the simulation system mass was updated to match the latest measurement of the Milky Way mass which is taken to be the upper limit of $\sim \SI{2.5e11}{M_{\odot}}$ determined by \citet{Jiao_2023}. An addition has been made to the source code that writes formation time, mass, number and relative probability of occurrence, according to the logarithmically binned IMF, into an output file, which allows tracking of the whole Pop III stellar population and further post-processing. In total, 100 simulations have been run with the same parameter set such that fluctuations in star formation emerging from differences in initial conditions could be mitigated. The Star Formation Rates (SFRs) from the individual simulations and their average are shown in Fig. ~\ref{figuire:ASLOTH_SFR}. The parameterization of the average, which will from now on be called the A-SLOTH SFR, is given by: 

\vspace{-0.2cm}

\begin{equation} \label{A}
    \psi_{A}(z) = \psi_{2}(z) + \psi_{4}(z),
\end{equation}

\vspace{0.3cm}

\noindent with 

\vspace{-0.2cm}

\begin{equation} \label{O2}
        \log_{10} \left(  \frac{\psi_{2}(z)}{\SI{}{M_{\odot} \, yr^{-1} \, Mpc^{-3}}} \right) = a + b \, z' + c \, z' \, ^{2},
\end{equation}

\vspace{-0.2cm}

\begin{equation} \label{O4}
        \log_{10} \left(  \frac{\psi_{4}(z)}{\SI{}{M_{\odot} \, yr^{-1} \, Mpc^{-3}}} \right) = a + b \, z' + c \, z' \, ^{2} + d \, z' \, ^{3} + e \, z' \, ^{4},
\end{equation}

\vspace{0.2cm}

\noindent where $z' = z + z_{0}$. The fit parameters of the parameterization are given in Table ~\ref{table:fit_params}. The parameterization is purposefully cut off at both high and low redshifts to account for the start of Pop III star formation and its termination. While the starting redshift results from the analytical prescriptions implemented in the A-SLOTH code, the termination redshift is implied from the absence of Pop III stars in our cosmic neighborhood. This indicates Pop III star formation to have effectively stopped. For this work, the formation of metal-free stars is thus assumed to occur until the redshift of the closest observed candidate object that provides indication of the existence of Pop III stars. This is the Lynx Arc. A strongly gravitationally-lensed structure at redshift $z = 3.357$ \citep{fosbury_massive_2003} \footnote{Currently, the galaxy CR3 at z = 3.193 provides closest indication for the existence of Pop III stars (\citet{Cai_2025}). The observation of CR3 has been reported after this work had been carried out. Using z = 3.193 as the termination of Pop III formation instead of the used value of z = 3.357 would have caused a negligible increase in IMBH numbers from Pop III stars.}.

\begin{figure}[h]
    \centering
    \includegraphics[scale = 0.35]{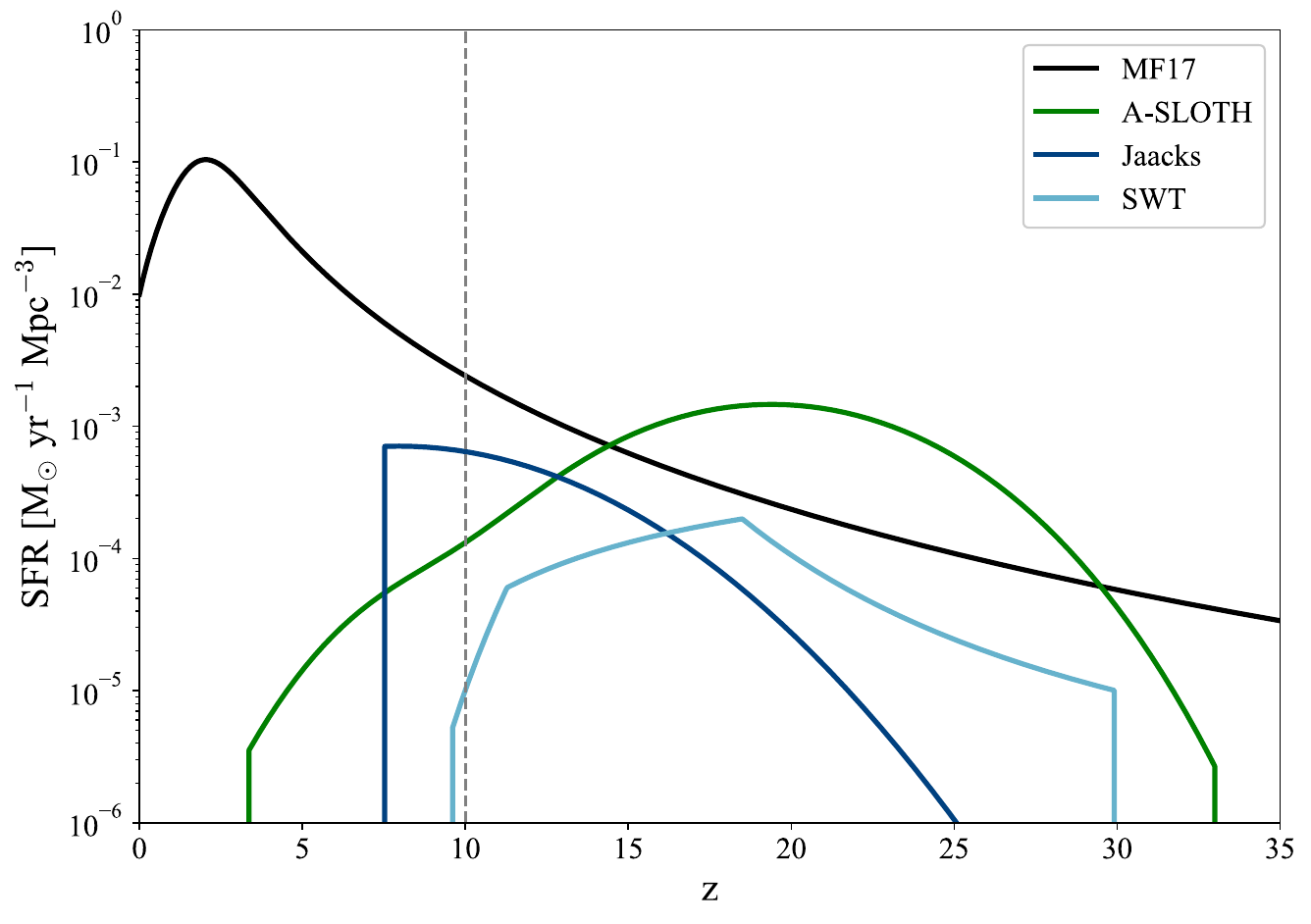}
    \caption{Comparison of Pop III A-SLOTH, Jaacks and SWT SFRs with MF17. A-SLOTH Pop III star formation is terminated at $z = 3.357$.}
    \label{figure:SFRs}
\end{figure}

For comparison, the A-SLOTH SFR is viewed along with two other Pop III SFRs in Fig. ~\ref{figure:SFRs}: The SFR presented by \citet{Jaacks_2019} results from a hydrodynamical simulation of the early universe that was halted at redshift $z \sim 7$. It is suggestive of robust ongoing Pop III star formation to be possible at $z < 7$ in the presence of Pop II stars. The parametrized Jaacks SFR follows: 

\vspace{-0.2cm}

\begin{equation} \label{equation:Jaacks}
\begin{split}
       & \log_{10} \left( \frac{\psi_{J}(z)}{\SI{}{\SI{}{M_{\odot} \, yr^{-1} \, Mpc^{-3}}}} \right) = \\ & - \SI{3.7626e+00}{} + \SI{1.5430e-01}{} \, z - \SI{9.7130e-03}{} \, z^2.
\end{split}
\end{equation}

\vspace{0.3cm}

The other Pop III SFR shown in Fig. ~\ref{figure:SFRs} is a simplified parametrization of the \cite{Skinner_2020} Pop III SFR that was adopted by \citet{Tanikawa_2022} in their study of reproducing the Gravitational Wave Transient Catalog 2 (GWTC-2) BBH merger rate. It shall be referred to as the SWT SFR. Shown in Fig. ~\ref{figure:SFRs} is also the best fit model of total star formation by \citet{Madau_2017} constructed from multi-wavelength measurements in the redshift range $z = 0 - 10$ that will be called the MF17 SFR.

\subsection{Remnants of Population III Stars} \label{section:remnants}
The evolution of stars and the type of remnant they leave behind is highly dependent on their mass. Zero-metallicity progenitor stars $\lesssim \SI{20}{M_{\odot}}$ are generally incapable of forming a BH remnant \citep{woosley_evolution_2002}. For stars above $\gtrsim \SI{20}{M_{\odot}}$ BH formation is possible either by fallback or direct collapse \citep{Fryer_2012}, where in the latter case the full stellar mass is assumed to be retained as metal-free stars do not suffer significant mass loss from intense winds \citep{Belczynski_2016}. Stars with helium cores above $\sim 30 - \SI{50}{M_{\odot}}$, experience a phenomenon known as pair instability \citep{Woosley_2017}, where photons produced in the core of the star are sufficiently energetic to produce electron-positron pairs via $\gamma \gamma \longrightarrow e^{+} e^{-}$. The subsequent loss of radiation pressure causes an explosive ignition of fusion to counteract the stellar collapse leading to a recurring Pulsational Pair-Instability Supernova (PPISN) or, if the first pulse is sufficiently energetic to disrupt the star, a Pair-Instability Supernova (PISN). A PPISN leaves behind a BH remnant with a mass given by the maximal stable helium-core mass, whereas a PISN does not leave behind a remnant at all.

\begin{figure}[H]
    \centering
    \includegraphics[scale = 0.35]{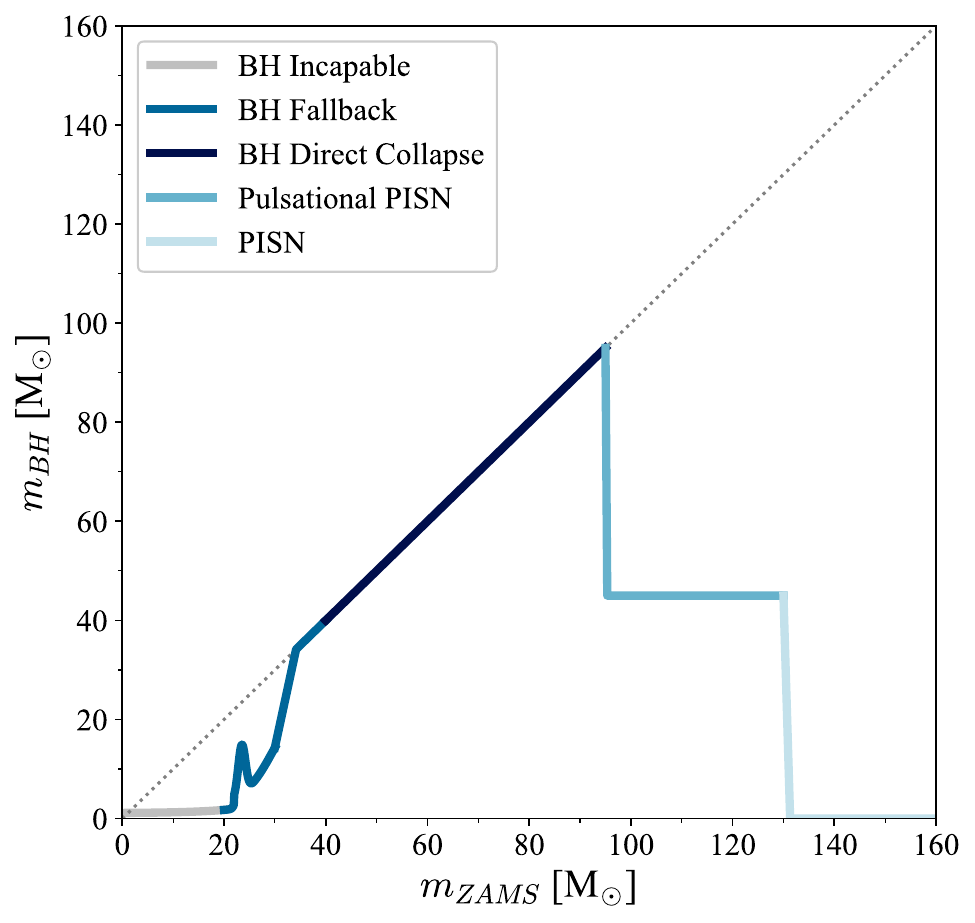}
    \caption{Remnant mass function of metal-free stars following \citet{Fryer_2012}. The mass range above $\SI{260}{M_{\odot}}$ is not shown, though it simply follows the dotted line (BH retains full progenitor mass). For a comparison of the stellar model with inefficient convective overshoot (this plot) with a model with efficient convective overshoot see Fig. 2 in \citet{Tanikawa_2021}.}
    \label{figure:w}
\end{figure}

\begin{figure}[H]
    \centering
    \includegraphics[scale = 0.35]{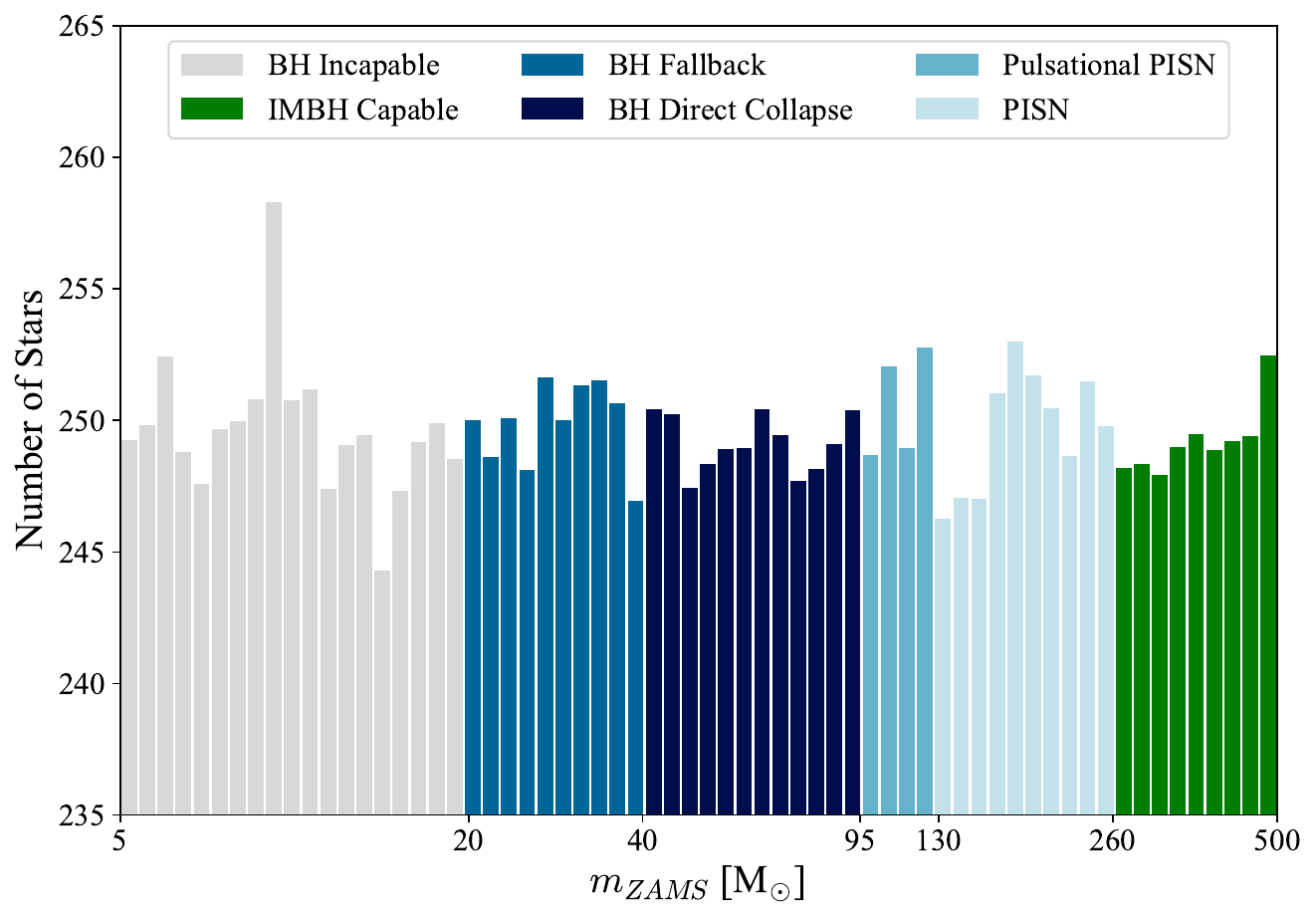}
    \caption{Averaged mass distribution of Pop III stars from 100 A-SLOTH simulations. The sampling reflects the log-flat IMF. Coloring indicates different stellar evolutionary paths depending on the ZAMS mass.}
    \label{figure:IMF}
\end{figure}

\noindent To create cores with $\sim 30 - \SI{50}{M_{\odot}}$, Zero-Age-Main-Sequence (ZAMS) masses range from $\sim \SI{70}{M_{\odot}}$ \citep{Woosley_2017} to $\sim \SI{95}{M_{\odot}}$ \citep{Tanikawa_2021}. In this work, the start of the pair instability is marked by a helium core of $\SI{45}{M_{\odot}}$, equivalent to the ZAMS mass of $\SI{95}{M_{\odot}}$. This model was assumed by \citet{Tanikawa_2022} to simulate the Pop III BBH merger rate that is used to determine the number of unmerged IMBHs. PISN occur for stars in the range $130 - \SI{260}{M_{\odot}}$. Beyond ZAMS masses of $\sim \SI{260}{M_{\odot}}$ photodisintegration of the collapsing core caused by pair instability prevents explosive fusion such that a direct collapse to a BH is possible again devouring the whole star \citep{Fryer_2001,Renzo_2020}. Furthermore, the term ``high-mass Pop III'' is reserved for masses beyond $\SI{260}{M_{\odot}}$ which are capable of forming IMBH remnants. The different evolutionary paths are indicated in Fig. ~\ref{figure:w} and Fig. ~\ref{figure:IMF} that show the remnant mass function and the sampling of Pop III stars by the A-SLOTH code respectively.

\subsection{Population III Binary Black Holes} \label{section:binaries}
The number of unmerged IMBHs follows from the assessment of BBH mergers that occur in a population of Pop III remnants. The nature and occurence of BBH mergers has been studied using gravitational wave observations \citep{Abbott_2022} which however is limited to lower-mass systems. Instead, this work refers to simulations by \citet{Tanikawa_2022} reproducing the BBH merger rate from GWTC-2 and extrapolates the merging behavior to higher masses. \citet{Tanikawa_2022} account for a great variety of stellar progenitor types including Pop III stars which allows to view mergers of Pop III BBHs in isolation. They find that the differential merger rate wrt. the primary mass above $m_{1} = \SI{45}{M_{\odot}}$ is dominated by Pop III remnants. With the simulated merger rate being in agreement with GWTC-2 at high primary masses, this puts a very strict constraint on the amount of allowed Pop III star formation. An SFR that's too large would overproduce the measured BBH merger rate at high primary masses and one is thus limited to the SWT SFR used by \cite{Tanikawa_2022} or an SFR with equivalent cumulative star formation. The number of Pop III stars simulated using the A-SLOTH code follows from the A-SLOTH SFR that well exceeds SWT in terms of cumulative star formation.

\noindent Since star formation is assumed to take place with a constant log-flat IMF throughout cosmic time for both the A-SLOTH and SWT SFRs, the A-SLOTH stellar population is simply down-scaled to match the population size that would be expected for the SWT SFR. The required scaling factor is determined by the ratio of the SFR integrals as given by:

\vspace{-0.1cm}

\begin{equation} \label{SWT_scaling}
    s_{SWT/A} = \frac{\int_{z_{min}}^{z_{max}} \psi_{SWT}(z) \, \mathrm{d}z}{\int_{z_{min}}^{z_{max}} \psi_{A}(z) \, \mathrm{d}z}.
\end{equation}

\subsubsection{Effective Volume Occupied by Stellar Population} \label{section:volume}
For the BBH merger rate, simulated by \citet{Tanikawa_2022}, to be applied to the population of remnant BHs from Pop III stars, a certain spatial understanding of the distribution of the initial stellar population is required. Due to A-SLOTH being unable to track the positions of the various objects throughout the simulation, spatial information on the stars in the Pop III catalog and their remnant BHs is unavailable. Thus, one must rely on the SFR to relate the population of stars to a volume in which merger events are expected to occur. Following \citet{madau_cosmic_2014} the SFR can be translated into a stellar remnant mass density or ``Dead'' Mass Density (DMD) $\rho_{\dag}$. The conversion is given by:


\vspace{-0.1cm}

\begin{equation} \label{DMD}
    \rho_{\dag}(z) = D \int_{z}^{z_{max}} \psi(z') \, \abs*{\frac{\mathrm{d}t'}{\mathrm{d}z'}} \, \mathrm{d}z'.
\end{equation}

\vspace{0.3cm}

\noindent The ``dead fraction'' $D$ can be understood as the fraction of mass of a stellar population that ends up as stellar remnants and is given by the following equation

\vspace{-0.1cm}

\begin{equation} \label{dead}
    D = \int\limits_{m_{min}}^{m_{max}} w(m) \, \xi(m) \, \mathrm{d}m,
\end{equation}

\vspace{0.3cm}

\noindent where $w(m)$ is the remnant mass function.


\begin{figure}[H]
    \centering
    \includegraphics[scale = 0.35]{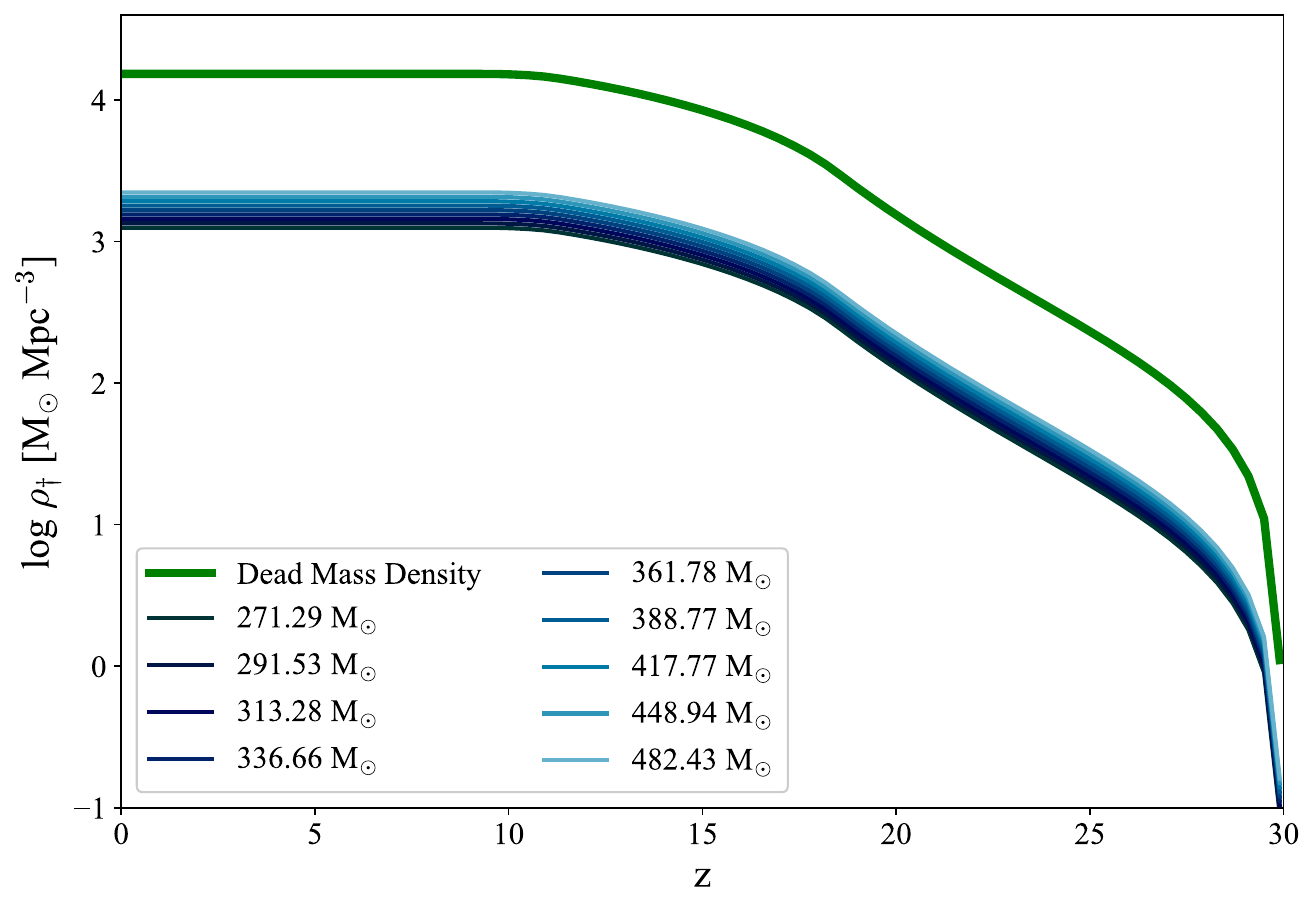}
    \caption{Total dead mass density of all IMBH remnants (green) and dead mass densities for the individual IMBH capable bins (blue lines).}
    \label{figure:DMDs}
\end{figure}


\noindent Making the restriction for masses in the calculation of Eq.~\eqref{dead} to include only high-mass Pop III stars ($> \SI{260}{M_{\odot}}$), Eq.~\eqref{DMD} yields the mass density of IMBHs. This IMBH DMD is further split to obtain individual DMDs for each A-SLOTH high-mass bin $i$ by weighting the IMBH DMD with the ratio of the sampling probability of each bin $p_{i}$ and the probability to sample a high-mass star $p_{HM}$ as given by:

\begin{equation} \label{DMD_bin}
    \rho_{\dag, \, i}(z) = \frac{p_{i}}{p_{HM}} \, \rho_{\dag}(z).
\end{equation}

\vspace{0.3cm}

\noindent The full IMBH DMD and the individual high-mass bin DMDs are illustrated in Fig. ~\ref{figure:DMDs}. The division of each individual DMD by its bin mass then converts it to the number density for each bin. The derived number densities at present day ($z = 0$) can finally be used to relate the population of stars, and subsequently their BH remnants, to an ``effective volume'' $V_{eff}$ which they would occupy under the assumption of a homogeneous spatial distribution of the objects:

\vspace{0.1cm}

\begin{equation} \label{V_eff}
    V_{eff, \, i} = s_{SWT/A} \frac{N_{i} \, m_{i}}{\rho_{\dag, \, i}(z = 0)},
\end{equation}

\vspace{0.3cm}

\noindent where $N_{i}$ is the number of objects in the high-mass bin $i$ and $m_{i}$ is the bin mass. The factor $s_{SWT/A}$ is used to scale the population size corresponding to the A-SLOTH SFR to match the SWT SFR. In view of the number densities, $V_{eff, \, i}$ is the same for all bins.

\subsubsection{Binary Black Hole Combinatorics} \label{section:combinatorics}
The BBH merger rate can be applied to the effective volume $V_{eff}$ to determine the number of mergers, given the extent of the initial stellar population. A comparison of the number of mergers with the initial population size reveals how many IMBHs remain unmerged. For this, the Pop III BBH merger rate $R_{T22}$ from Fig. 1 of \citet{Tanikawa_2022} is integrated over redshift, as given by: 

\vspace{-0.2cm}

\begin{equation} \label{MND}
    R_{merge} = \int\limits_{z = 0}^{z_{max}} R_{T22}(z') \abs*{\frac{\mathrm{d}t'}{\mathrm{d}z'}} \, \mathrm{d}z'.
\end{equation}

\vspace{0.3cm}

\noindent One obtains the number of mergers per unit volume which amounts to $R_{merge} = \SI{17.72}{Mpc^{-3}}$. This merger number density contains events with all the various pairings of primary and secondary BH masses that are possible. In order to make a statement on the merger number density for events specifically containing IMBHs, an overview of all different BBH mass combinations and their relative likelihoods of occurrence is necessary.

\begin{figure}
    \centering
    \includegraphics[scale = 0.37]{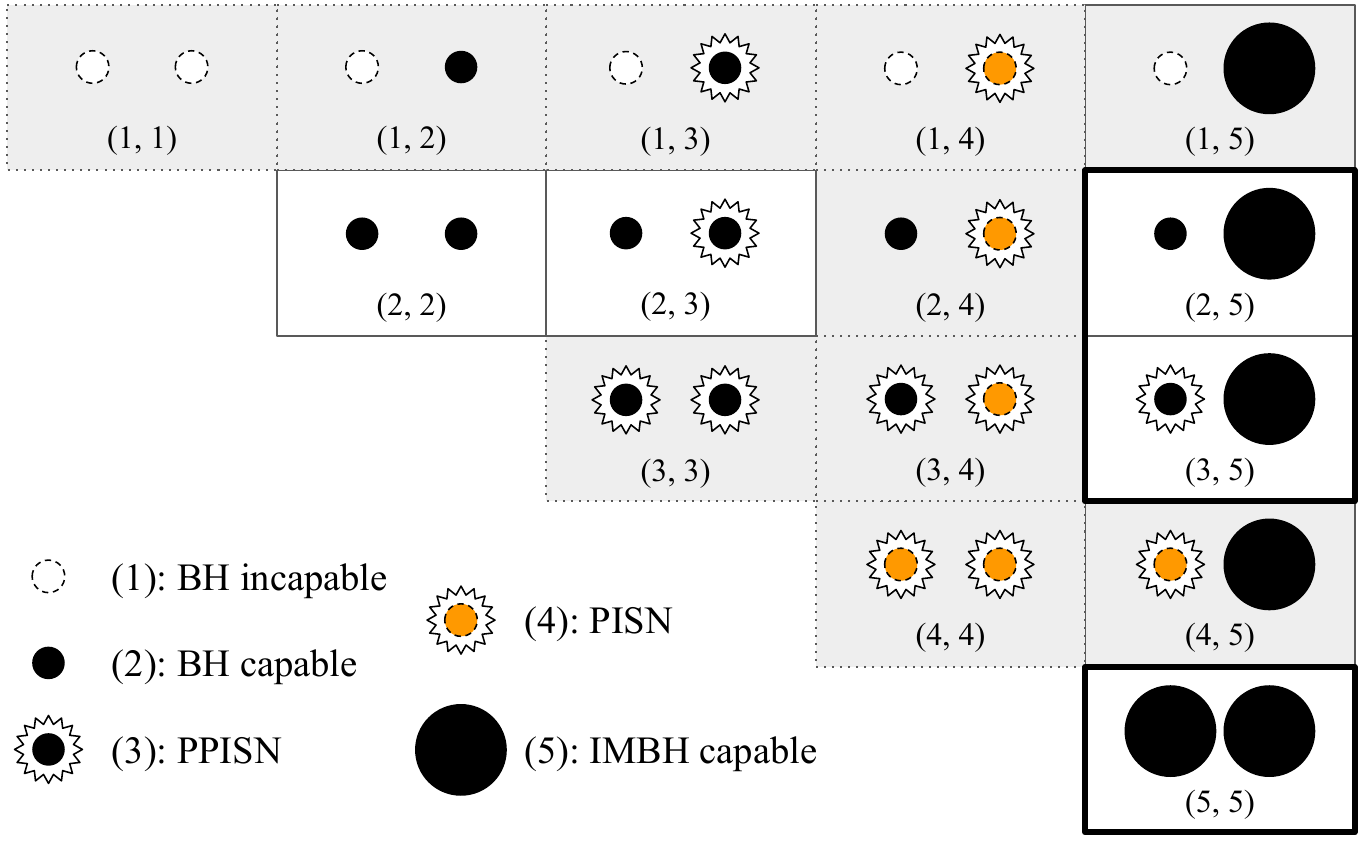}
    \caption{Combinations of binary stellar systems shown for 5 mass bins. Combinations inside dotted outline simulated in \cite{Tanikawa_2022}. Combinations inside thin solid outline lead to BBH mergers in \cite{Tanikawa_2022}. Combinations inside thick solid outline lead to mergers of IMBHs (this work).}
    \label{combinatorics}
\end{figure}

\citet{Tanikawa_2022} operate in the scenario where BBH systems arise as a consequence of evolved Pop III stellar binary systems. It follows, that the frequency of occurrences is given by the initial stellar pairings where the BH masses in question are determined via the remnant mass function. 
Naturally, the formation of a Pop III binary with a particular pairing of masses results from the complicated physical processes governing the fragmentation of their common protostellar disc. Since studying this demands extensive and resource intensive N-body simulations as performed by \citet{stacy_constraining_2013} and \citet{stacy_building_2016}, a simplified approach is chosen where the occurrence of a mass pairing is solely given by the occurrence of the single stars themselves. 
One might picture an urn filled with stars of different masses from which two stars are picked at random to comprise a binary system. For our purposes, this urn is to be identified with the population of stars as tracked by A-SLOTH. Since all $n_{bin}$ mass bins are sampled virtually equally, one ends up with $n_{bin}^{2}$ combinations, all sharing the same relative occurrence. Fig. ~\ref{combinatorics} illustrates the combinatorics for a simplified case with one bin for every type of stellar evolution (1. BH incapable, 2. BH capable, 3. PPISN, 4. PISN, 5. IMBH capable). Note, that only a subset of the 64 mass bins in A-SLOTH are considered for this calculation. This is for the following two reasons: 
\begin{enumerate}
    \item  Certain initial stellar binaries do not lead to mergers. For one, Pop III stars with masses $< \SI{20}{M_{\odot}}$ are too light to form BHs in the first place. Hence, all bins with $m_{i} < \SI{20}{M_{\odot}}$ are excluded from the calculation (all binary combinations containing 1-type stars in Fig. ~\ref{combinatorics} are excluded). Effects of pair-instability demand another augmentation of the number of possible combinations. Stars ending their lives in a PISN do not leave any remnants behind.
As such, all combinations, where at least one of the binary partners has a mass $\SI{130}{M_{\odot}} \leq m_{i} \leq \SI{260}{M_{\odot}}$, do not
contribute (all binary combinations containing 4-type stars in Fig. ~\ref{combinatorics} are excluded). Finally, systems in which both stars undergo PPISN and leave behind a $\SI{45}{M_{\odot}}$ BH also don't contribute (combination (3, 3) in Fig. ~\ref{combinatorics}). This is due to the immense mass loss stars can experience during a PPISN. Mass loss can significantly shrink the stars and thus eliminate the possibility of stellar interaction, which under normal circumstances would accelerate the rate of inspiral or disrupt the binary altogether. All $n_{PPISN}$ bins with $n_{PPISN}^{2}$ pairings of PPISN-capable stars are subtracted from the list of merging combinations. 
\item  \citet{Tanikawa_2022} has defined the upper mass limit of Pop III stars to be $\SI{150}{M_{\odot}}$, while this work includes stars with masses up to $\SI{500}{M_{\odot}}$. Considering the effects of pair instability, as discussed in Sect. ~\ref{section:remnants}, IMBHs cannot be formed in \cite{Tanikawa_2022} at all. This implies that the mergers per unit volume of around $\SI{17.72}{Mpc^{-3}}$ do not account for the mergers of IMBHs. Rather, it can be understood as a reduced number of mergers corresponding to objects $< \SI{150}{M_{\odot}}$ (combinations (2, 2) and (2, 3)in Fig. ~\ref{combinatorics}).
\end{enumerate}

\noindent Statements about the mergers of IMBHs can be made by exploiting the equal relative occurrence of all initial stellar mass combinations. If the merger number density for a single pairing is determined, it can be multiplied with the number of combinations containing IMBH progenitors to yield the number of mergers missing in \cite{Tanikawa_2022}. Following Fig. ~\ref{combinatorics} and the augmentations to the contributing BBH combinations to the mergers in \cite{Tanikawa_2022}, the number of relevant initial stellar bins equates to $n_{bin} = 26$, with $n_{PPISN} = 4$ being the number of bins that yield $\SI{45}{M_{\odot}}$ remnants. The resulting number of contributing combinations to the cumulative mergers per unit volume is:

\vspace{-0.2cm}

\begin{equation}
    n_{comb} = n_{bin}^{2} - n_{PPISN}^{2} = 660.
\end{equation}

\vspace{0.3cm}

\noindent The individual merger number density for any one of the 660 combinations $R_1$ naturally is:

\vspace{-0.1cm}

\begin{equation} \label{R1}
    R_1 = \frac{R_{merge}}{n_{comb}} \approx \SI{0.03}{Mpc^{-3}}.
\end{equation}

\vspace{0.3cm}

\noindent One can compute the number of objects per bin ``lost'' in the mergers throughout cosmic time by multiplying $R_1$ with the object ``loss multiplicity'' i.e., the aggregated number of objects from a bin lost in the various combinations due to mergers. Generalizing the simplified scheme shown in Fig ~\ref{combinatorics} and extending the number of bins to $n_{bin, \, HM} = 35$ to include the high-mass stars, in other words the IMBHs, which were previously missing in $n_{bin}$, the loss multiplicity $N_{loss}$ is given by:

\vspace{-0.1cm}

\begin{equation} \label{multiplicity}
    N_{loss} = 2 (n_{bin, \, HM} - 1) + 2 = 2n_{bin, \, HM}
\end{equation}

\vspace{0.4cm}

\noindent The first term of the sum represents the number of objects lost from combinations where the binary partners have different masses. The factor 2 is included to account for the two combinatorial pairings that yield the same physical binary (e.g. pairings (5, 3)  and (3, 5), while combinatorially unique with equal relative occurrence, both yield a physical system (5, 3) with the combined relative occurrence). The second term of the sum accounts for the pairing of objects with identical masses (the case (5, 5)). In this instance, a merger event will reduce the number of objects in the bin by 2.
Combining Eqs. ~\eqref{V_eff}, ~\eqref{R1} and ~\eqref{multiplicity}, the number of merged objects per bin can be obtained and is given by:

\vspace{-0.2cm}

\begin{equation} \label{N_merged}
    N_{merged, \, i} = R_1  \, N_{loss} \, V_{eff, \, i}.
\end{equation}

\vspace{0.3cm}

\noindent A sum of Eq.~\eqref{N_merged} over all high-mass bins returns the number of IMBHs that have been lost to mergers throughout cosmic time: 

\begin{equation} \label{N_merged_sum}
    N_{merged} = \sum_{i, \,\,\,  m_{i} \geq 260} N_{merged, \, i}.
\end{equation}

\vspace{0.3cm}

\noindent Naturally, the difference between the initial number of IMBHs $N_{HM}$ and $N_{merged}$ gives the number of IMBHs that have remained unmerged until present day: 

\vspace{-0.2cm}

\begin{equation}
    N_{U} = N_{HM} - N_{merged}.
\end{equation}

\section{Results} \label{section:results}

\subsection{Number of Unmerged IMBHs} \label{section:number}
The A-SLOTH simulation yields an average number of high-mass Pop III stars $\langle N\rangle = 2242.85$ with 
a statistical uncertainty of $\sigma = 103.55$ per Milky Way-like galaxy arising from 100 simulation runs.
In view of the stellar evolutionary paths of high-mass Pop III stars, as discussed in Sect. ~\ref{section:remnants}, that number is immediately to be identified as the population size of their remnant IMBHs. Compared to the A-SLOTH SFR, SWT yields about a tenth of cumulative star formation following Eq. ~\eqref{SWT_scaling} such that the IMBH population reduces to $N^{SWT} = 222.62 \pm 10.28$. This shall be regarded as the total number of IMBHs formed over cosmic time before mergers are taken into account. 

\begin{figure}[H]
    \centering
    \includegraphics[scale = 0.4]{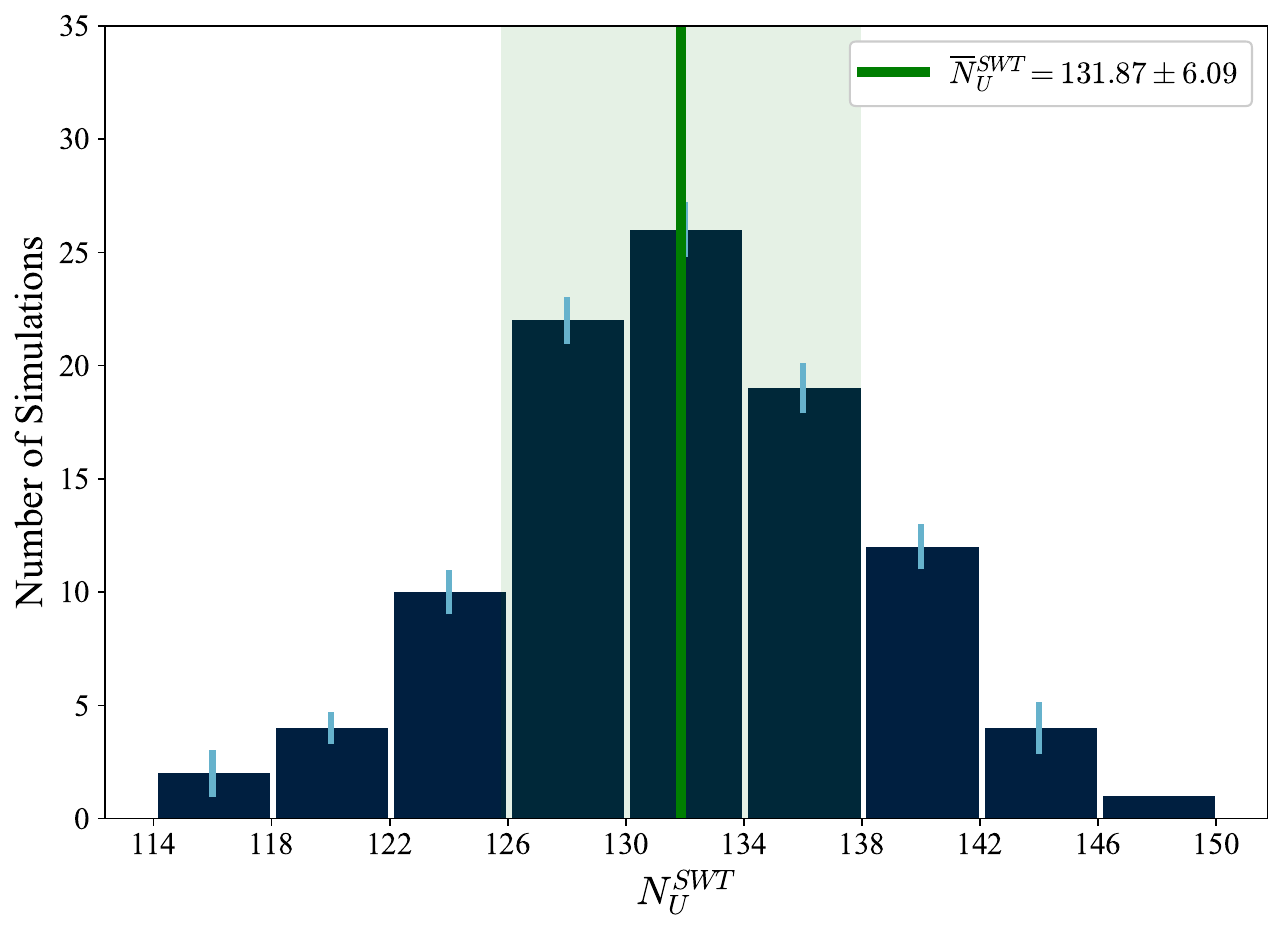}
    \caption{Distribution of the number of unmerged IMBHs across the 100 A-SLOTH simulations. for the SWT SFR. Light blue lines indicate the standard deviation for each bin.}
    \label{counts}
\end{figure}

The number of unmerged IMBHs, as inferred from \citet{Tanikawa_2022} merger rate simulations using the combinatorial assessment of BBH mergers in an effective volume, equates to $N_{U}^{SWT} = 131.87 \pm 6.09$. The uncertainty again denotes the statistical 1$\sigma$ spread about the mean value. Fig. ~\ref{counts} illustrates the distribution of the number of unmerged IMBHs across the 100 A-SLOTH simulations. 

Systematic uncertainties on the number of unmerged IMBHs are most likely greater than the variability resulting from 100 simulation runs. With the population size being sensitive to the underlying SFR, it is the primary source of systematic uncertainty in this work. Observations with gravitational wave detectors provide the most stringent constraints on the SFR yet. By reproducing high-mass BBH mergers using Pop III remnants, the result from this work is to be understood as an upper bound. Any SFR that yields a size in population below the one presented here cannot be excluded.

\subsection{Pop III Contribution to Extragalactic Background Light} \label{section:EBL}
In view of an IMBH forming population of Pop III stars constrained by observations of BBH mergers, it is investigated to what extent its light contributes to the EBL.
The extent of their contribution is determined following the equation for the specific emissivity $\epsilon_\nu$ presented by \citet{kneiske_implications_2002}:

\begin{equation} \label{emiss}
    \epsilon_{\nu}(z) = \int\limits_{z}^{z_{max}} L_{\nu}(t(z) - t(z')) \, \psi(z') \, \abs*{\frac{\mathrm{d}t'}{\mathrm{d}z'}} \, \mathrm{d}z'
\end{equation}

\vspace{0.3cm}

\noindent and the resulting spectral energy distribution of the EBL

\begin{equation} \label{ebl}
    P_{\nu}(z) = \nu I_{\nu}(z) = \nu \frac{c}{4 \pi} \int\limits_{z}^{z_{max}} \epsilon_{\nu'}(z') \, \abs*{\frac{\mathrm{d}t'}{\mathrm{d}z'}} \mathrm{d}z'.
\end{equation}

\vspace{0.3cm}

\noindent The quantities $L_{\nu}(t(z))$ and $\psi(z)$ are the Pop III luminosity and the SFR, respectively. The cosmological parameters enter via:

\begin{equation} \label{derivative}
    \abs*{\frac{dt}{dz}} = \frac{1}{H_{0} (1 + z) \, E(z)},
\end{equation}

\vspace{0.3cm}

\noindent with


\begin{equation} \label{cosmo}
    E(z) = \sqrt{\Omega_{\Lambda} + \Omega_{k} (1 + z)^{2} + \Omega_{m} (1 + z)^{3} + \Omega_{r} (1 + z)^{4}}.
\end{equation}

\begin{figure}[h]
    \centering
    \includegraphics[scale = 0.35]{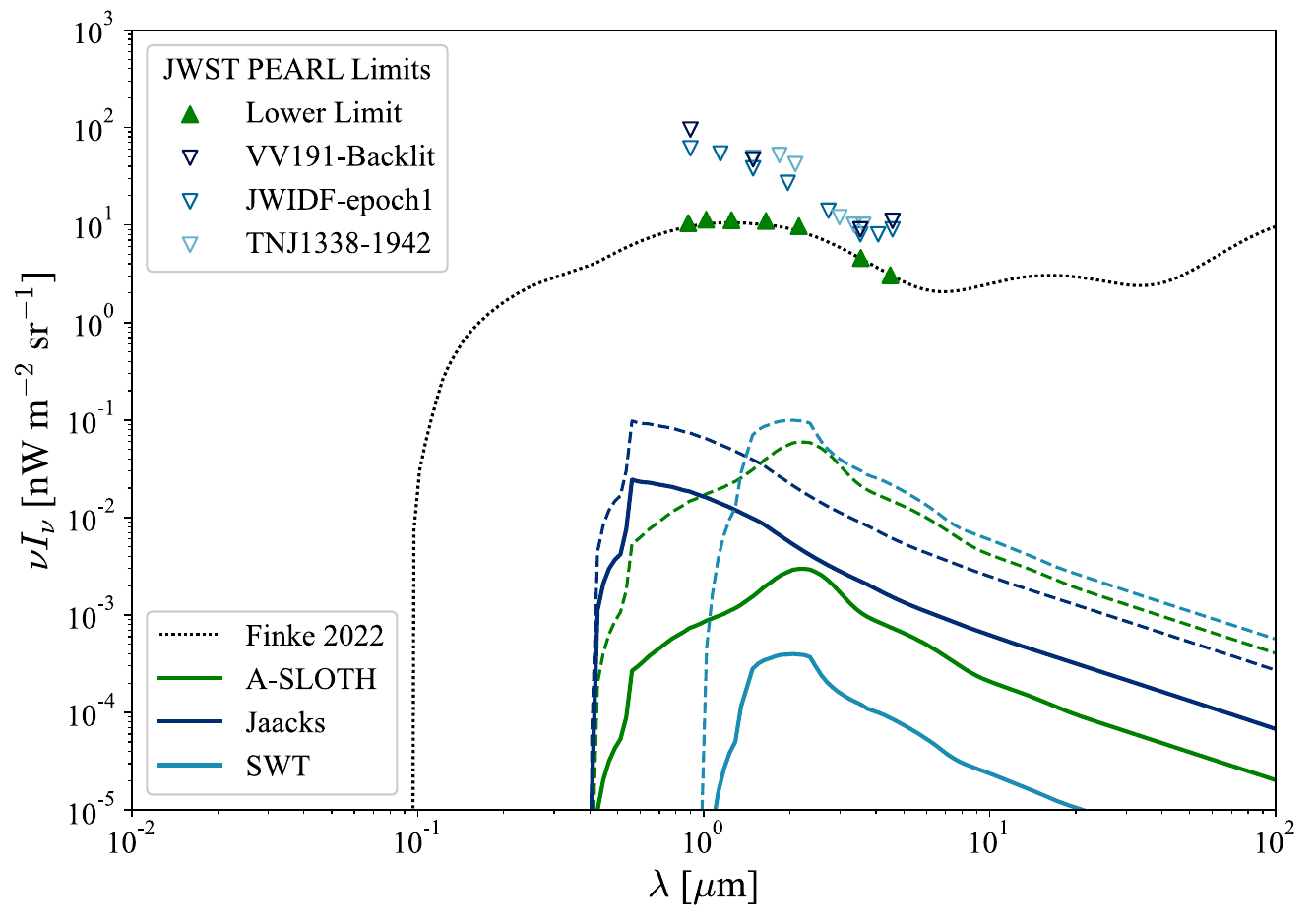}
    \caption{EBLs for configurations 1-A-0, 1-J-0 and 1-SWT-0 both for unscaled SFRs (solid lines) and scaled SFRs (dashed lines) with factors 20, 4 and 250 for A-SLOTH, Jaacks and SWT respectively. For comparison, the most recent model of EBL lower limits by \cite{Finke_2022} and JWST galaxy count and direct measurements (\cite{Windhorst_2022}).}
    \label{figure:1-AJSWT-0}
\end{figure}

\vspace{0.3cm}

\noindent where a flat $\Lambda$CDM cosmology with parameters from the \citet{planck_collaboration_planck_2016} was assumed. The intrinsic Pop III spectra and the reprocessed spectra from the ISM were modeled using fitting functions presented in \citet{Fernandez_2006}. The resulting Pop III contribution to the EBL is shown in Fig. ~\ref{figure:1-AJSWT-0} for the A-SLOTH, Jaacks and SWT SFRs. To assess the Pop III contribution to the EBL for the full star-forming period, the Jaacks SFR has been extended for $z < 7$. It has been extrapolated following Eq. ~\eqref{equation:Jaacks} to meet the redshift of the Lynx Arc at $z = 3.357$, which is assumed to be the time of Pop III star formation termination. Comparing the solid lines with EBL measurements by JWST \citep{Windhorst_2022} shows that the Pop III contribution to the EBL is, at best, at the $0.1\%$ level. In the case of the SWT SFR it is safe to assume that the contribution is negligible. While this would lead to the overproduction of BBH mergers, scaling the Pop III SFRs such that they match the farthermost available data of the measured MF17 SFR at redshift $z = 10$ raises the Pop III EBL contributions to the $1\%$ level. The scalings are 20 for A-SLOTH, 4 for Jaacks and 250 for SWT. While the calculation of the EBL from only Milky Way-like galaxies, as would be the case for the A-SLOTH SFR, would certainly introduce a bias, the above statement likely holds true for a generalized universe, since the numerical simulations of the Jaacks and SWT SFRs did not constrain the size and shape of the formed galaxies. It can be concluded that Pop III stars are, in effect, hidden in both the amount of star formation as well as the EBL. The number of unmerged IMBHs from Pop III stars determined in this work is, therefore, not in conflict with measured stellar observables. This result is also in agreement with \citet{Finke_2022}, who have presented a model of the optical and infrared bands of the EBL. While no explicit comments were made on the impact of Pop III stars, the contributions of light from different cosmic times to the present-day EBL suggest that the most significant impact comes from light emitted in the local universe. In fact, light from redshift $z \geq 3$ does not seem to play a role at all. A similar conclusion was made by \citep{Sun_2021} regarding the mean near infrared radiation background (NIRB) when employing a Pop III SFR model similar to the scaled Jaacks SFR in this work. Their modeling of the angular EBL fluctuations provides promising prospects on discerning the Pop III contribution, as the angular fluctuations show strong wavelength dependence for high multipoles due to prominent Lyman-$\alpha$ emission, which further increases with growing stellar mass. In the scenario suggested here, the expected EBL contribution is predominantly limited by the constraints on the SFR obtained through gravitational wave measurements of the merger rate. A detection of a signature of Pop III stars in the EBL appears to be unlikely given these constraints.

\vspace{-0.3cm}

\section{Discussion} \label{section:discussion}

\subsection{Caveats} \label{section:caveats}
Given the physical complexity of BBH mergers, as they emerge from Pop III stars, the method presented in this work provides a simplified approach to estimate the fraction of BBH merged systems.  An effort was made to account for initial pairings that would not result in BBH mergers, however, the assessment of individual binary systems was not possible. 
Rather, combinations were either categorically included or excluded using robust arguments based upon our current understanding of BBH formation.   
With all of this in mind, the uncertainty of the number of unmerged IMBHs with this method is bound to be considerably larger than the statistical $1\sigma$ spread among the 100 Milky Way-like galaxy simulations. 

With these shortcomings in mind, how reasonable is the calculated number of $N_{U}^{SWT} = 131.87 \pm 6.09$ unmerged IMBHs in a Milky Way-like galaxy? By constructing binary progenitor Pop III systems with a binary fraction of $f_B = 0.5$, the same one used by \citet{Tanikawa_2022} in their simulations, and making the naive assumption that all binary remnant systems are guaranteed to merge irrespective of stellar evolution and orbital dynamics, it is possible to determine a lower limit for the number of unmerged IMBHs. This number corresponds to the population size of solitary high-mass Pop III stars which, following a binary fraction of $f_B = 0.5$, is 1/3 of all high-mass objects. Therefore the lower limit is $N_{U, \, lower}^{\, SWT} = 74.21 \pm 3.43$. The number of unmerged IMBHs calculated with the method presented in this work thus neatly falls in between the total number of high-mass objects and the lower limit of unmerged objects. 

\subsection{Other IMBH formation Scenarios} \label{section:scenarios}
Unmerged IMBHs from Pop III stars present only a subset of objects that might be able to acquire a DM spike. The presented method, for example, does not account for giant stars that form following a common envelope stage of a binary stellar system. These do not create BBH merger signatures as the remnant of a giant star is just a solitary IMBH, even though the initial system was a stellar binary. This IMBH might then go on to acquire a DM spike. On the other hand, a second-generation IMBH that is produced in a merger might also dress with a DM spike if the merger happens sufficiently early to allow the DM spike to grow. Though likely, these mergers would not be picked up by gravitational wave detectors as they would happen too far away. In this scenario, dressed IMBHs can also be produced from the merger of stellar BHs. Additionally, primordial black holes may contribute to the distribution of IMBHs \citep{2016JCAP...11..036B,2024arXiv240716373P}. 

\subsection{Comparison With Other Works} \label{section:comparison}
Compared with previous work conducted on determining the number of unmerged IMBHs from Pop III stars, this project yields comparatively few objects. \cite{Bertone_2005} have used a method of populating simulated halos at redshift $z = 18$ with $\SI{100}{M_{\odot}}$ BHs if the smoothed primordial density field featured a 3$\sigma$ peak, resulting in a number of $1027 \pm 84$ unmerged IMBHs. While stellar observables like the SFR and the EBL would allow for a greater population of Pop III IMBHs, advances in gravitational wave astronomy put very strict constraints on the amount of allowed star formation and, subsequently, the IMBH population size. 

Complementary to the presented work is the assessment of ``heavy seed'' IMBH populations that form from the direct collapse of giant pristine gas clouds. A recent examination of the matter was conducted by \cite{aschersleben2024gamma}, where the results of the large-scale hydrodynamical simulation EAGLE (\cite{Schaye_2014}) were analyzed to create a mock catalog of unmerged direct collapse IMBHs in Milky Way-like galaxies. The seed BHs are assumed to form with a mass of $10^{5}$ M$_{\odot}$ in hosting galaxies which exceed a mass of $10^{10}$ M$_{\odot}/h$. 
This study reveals that, on average, a Milky Way-like galaxy and its respective satellites  host  $15_{-6}^{+9}$ ``heavy seed'' unmerged IMBHs; an order of magnitude less than the ``light seed'' IMBHs that form as Pop III remnants. 
A comparison with the $101 \pm 22$ ``heavy seeds'' determined by \cite{Bertone_2005} using a similar method as discussed above once again shows that previous analyzes had overestimated the number of IMBHs and that the latest experimental observations and advances in numerical simulations present a more reserved estimate. Even though ``heavy seeds'' appear to be 10 times less common than the ``light seed'' Pop III remnants, ``heavy seeds'' are expected to have far bigger DM spikes, potentially being sufficiently luminous to be spotted by current and future gamma-ray observatories making them ideal for ``smoking gun'' detections of DM.

And while present-day gravitational wave detectors have limited capabilities in making observations of IMBH mergers, the next-generation ground and space-based detectors will open up the intermediate mass range such that more stringent limits on both ``light seed'' as well as ``heavy seed'' populations and their formation mechanisms will be set (\cite{Fragione_2023}).

\begin{acknowledgements}
We thank Tilman Hartwig and Simon Glover for their assistance in operating the A-SLOTH simulation code. We also thank Justin Finke for providing information on methods and results regarding the contribution of early-type stars to the EBL. Many thanks are expressed towards Sara Porras Bedmar, whose help in performing EBL calculations is greatly appreciated. Special thanks are given to Ataru Tanikawa for extensive discussions on stellar evolutionary models and the merging behavior of stellar remnants which make up the foundation of this work. DH acknowledges support by the Deutsche Forschungsgemeinschaft (DFG, German Research Foundation) under Germany’s Excellence Strategy – EXC 2121 „Quantum Universe“ – 390833306.
\end{acknowledgements}

\bibliographystyle{aa}
\bibliography{references}

@article{Chan_2023,
	doi = {10.3847/2041-8213/acaafa},
	url = {https://doi.org/10.3847%2F2041-8213%2Facaafa},
	year = 2023,
	month = {jan},
	publisher = {American Astronomical Society},
	volume = {943},
	number = {2},
	pages = {L11},
	author = {Man Ho Chan and Chak Man Lee},
	title = {Indirect Evidence for Dark Matter Density Spikes around Stellar-mass Black Holes},
	journal = {The Astrophysical Journal Letters}
}

@article{Aharonian_2008,
	doi = {10.1103/physrevd.78.072008},
	url = {https://doi.org/10.1103%2Fphysrevd.78.072008},
	year = 2008,
	month = {oct},
	publisher = {American Physical Society ({APS})},
	volume = {78},
	number = {7},
	author = {{HESS Collaboration}},
	title = {Search for gamma rays from dark matter annihilations around intermediate mass black holes with the {HESS} experiment},
	journal = {Physical Review D}
}

@article{Bertone_2005,
	doi = {10.1103/physrevd.72.103517},
	url = {https://doi.org/10.1103%2Fphysrevd.72.103517},
	year = 2005,
	month = {nov},
	publisher = {American Physical Society ({APS})},
	volume = {72},
	number = {10},
	author = {Gianfranco Bertone and Andrew R. Zentner and Joseph Silk},
	title = {New signature of dark matter annihilations: Gamma rays from intermediate-mass black holes},
	journal = {Physical Review D}
}

@article{Schaye_2014,
   title={The EAGLE project: simulating the evolution and assembly of galaxies and their environments},
   volume={446},
   ISSN={0035-8711},
   url={http://dx.doi.org/10.1093/mnras/stu2058},
   DOI={10.1093/mnras/stu2058},
   number={1},
   journal={Monthly Notices of the Royal Astronomical Society},
   publisher={Oxford University Press (OUP)},
   author={Schaye, Joop and Crain, Robert A. and Bower, Richard G. and Furlong, Michelle and Schaller, Matthieu and Theuns, Tom and Dalla Vecchia, Claudio and Frenk, Carlos S. and McCarthy, I. G. and Helly, John C. and Jenkins, Adrian and Rosas-Guevara, Y. M. and White, Simon D. M. and Baes, Maarten and Booth, C. M. and Camps, Peter and Navarro, Julio F. and Qu, Yan and Rahmati, Alireza and Sawala, Till and Thomas, Peter A. and Trayford, James},
   year={2014},
   month=nov, pages={521–554} }

@article{Fernandez_2006,
  title={The cosmic near-infrared background: Remnant light from early stars},
  author={Fernandez, Elizabeth R and Komatsu, Eiichiro},
  journal={The Astrophysical Journal},
  volume={646},
  number={2},
  pages={703},
  year={2006},
  publisher={IOP Publishing}
}

@article{Finke_2022,
  title={Modeling the Extragalactic Background Light and the Cosmic Star Formation History},
  author={Finke, Justin D and Ajello, Marco and Dom{\'\i}nguez, Alberto and Desai, Abhishek and Hartmann, Dieter H and Paliya, Vaidehi S and Saldana-Lopez, Alberto},
  journal={The Astrophysical Journal},
  volume={941},
  number={1},
  pages={33},
  year={2022},
  publisher={IOP Publishing}
}

@article{woosley_evolution_2002,
	title = {The evolution and explosion of massive stars},
	volume = {74},
	issn = {0034-6861, 1539-0756},
	url = {https://link.aps.org/doi/10.1103/RevModPhys.74.1015},
	doi = {10.1103/RevModPhys.74.1015},
	language = {en},
	number = {4},
	urldate = {2024-02-01},
	journal = {Reviews of Modern Physics},
	author = {Woosley, S. E. and Heger, A. and Weaver, T. A.},
	month = nov,
	year = {2002},
	pages = {1015--1071},
}

@article{stacy_constraining_2013,
	title = {Constraining the {Statistics} of {Population} {III} {Binaries}},
	volume = {433},
	issn = {1365-2966, 0035-8711},
	url = {http://arxiv.org/abs/1211.1889},
	doi = {10.1093/mnras/stt789},
	language = {en},
	number = {2},
	urldate = {2024-01-16},
	journal = {Monthly Notices of the Royal Astronomical Society},
	author = {Stacy, Athena and Bromm, Volker},
	month = aug,
	year = {2013},
	note = {arXiv:1211.1889 [astro-ph]},
	pages = {1094--1107},
}

@article{stacy_building_2016,
	title = {Building up the {Population} {III} initial mass function from cosmological initial conditions},
	volume = {462},
	issn = {0035-8711, 1365-2966},
	url = {http://arxiv.org/abs/1603.09475},
	doi = {10.1093/mnras/stw1728},
	language = {en},
	number = {2},
	urldate = {2024-01-16},
	journal = {Monthly Notices of the Royal Astronomical Society},
	author = {Stacy, Athena and Bromm, Volker and Lee, Aaron T.},
	month = oct,
	year = {2016},
	note = {arXiv:1603.09475 [astro-ph]},
	pages = {1307--1328},
}

@article{madau_cosmic_2014,
	title = {Cosmic {Star} {Formation} {History}},
	volume = {52},
	issn = {0066-4146, 1545-4282},
	url = {http://arxiv.org/abs/1403.0007},
	doi = {10.1146/annurev-astro-081811-125615},
	language = {en},
	number = {1},
	urldate = {2024-01-30},
	journal = {Annual Review of Astronomy and Astrophysics},
	author = {Madau, Piero and Dickinson, Mark},
	month = aug,
	year = {2014},
	note = {arXiv:1403.0007 [astro-ph]},
	pages = {415--486},
}

@article{fosbury_massive_2003,
	title = {Massive star formation in a gravitationally-lensed {HII}-galaxy at z=3.357},
	volume = {596},
	issn = {0004-637X, 1538-4357},
	url = {http://arxiv.org/abs/astro-ph/0307162},
	doi = {10.1086/378228},
	language = {en},
	number = {2},
	urldate = {2024-02-16},
	journal = {The Astrophysical Journal},
	author = {Fosbury, R. A. E. and Villar-Martin, M. and Lombardi, M. and Rosati, P. and Stern, D. and Hook, R. N. and Holden, B. P. and Stanford, S. A. and Squires, G. K. and Rauch, M. and Sargent, W. L. W.},
	month = oct,
	year = {2003},
	note = {arXiv:astro-ph/0307162},
	pages = {797--809},
}

@article{kneiske_implications_2002,
	title = {Implications of {Cosmological} {Gamma}-{Ray} {Absorption} - {I}.{Evolution} of the {Metagalactic} {Radiation} {Field}},
	volume = {386},
	issn = {0004-6361, 1432-0746},
	url = {http://arxiv.org/abs/astro-ph/0202104},
	doi = {10.1051/0004-6361:20020211},
	language = {en},
	number = {1},
	urldate = {2023-10-18},
	journal = {Astronomy \& Astrophysics},
	author = {Kneiske, Tanja M. and Mannheim, Karl and Hartmann, Dieter H.},
	month = apr,
	year = {2002},
	note = {arXiv:astro-ph/0202104},
	pages = {1--11},
}

@article{planck_collaboration_planck_2016,
	title = {\textit{{Planck}} 2015 results: {XIII}. {Cosmological} parameters},
	volume = {594},
	issn = {0004-6361, 1432-0746},
	shorttitle = {\textit{{Planck}} 2015 results},
	url = {http://www.aanda.org/10.1051/0004-6361/201525830},
	doi = {10.1051/0004-6361/201525830},
	language = {en},
	urldate = {2024-03-04},
	journal = {Astronomy \& Astrophysics},
	author = {{Planck Collaboration}},
	month = oct,
	year = {2016},
	pages = {A13},
}

@article{Noyola_2008,
   title={Gemini and Hubble Space TelescopeEvidence for an Intermediate‐Mass Black Hole in $\omega$ Centauri},
   volume={676},
   ISSN={1538-4357},
   url={http://dx.doi.org/10.1086/529002},
   DOI={10.1086/529002},
   number={2},
   journal={The Astrophysical Journal},
   publisher={American Astronomical Society},
   author={Noyola, Eva and Gebhardt, Karl and Bergmann, Marcel},
   year={2008},
   month=apr, pages={1008–1015} 
}

@article{1976MNRAS.176..633F,
    author = {Frank, Juhan and Rees, Martin J.},
    title = "{Effects of Massive Central Black Holes on Dense Stellar Systems}",
    journal = {Monthly Notices of the Royal Astronomical Society},
    volume = {176},
    number = {3},
    pages = {633-647},
    year = {1976},
    month = {09},
    issn = {0035-8711},
    doi = {10.1093/mnras/176.3.633},
    url = {https://doi.org/10.1093/mnras/176.3.633},
    eprint = {https://academic.oup.com/mnras/article-pdf/176/3/633/9402962/mnras176-0633.pdf},
}

@article{2001ApJ...552L.109K,
   title={Ultraluminous X-Ray Sources in External Galaxies},
   volume={552},
   ISSN={0004-637X},
   url={http://dx.doi.org/10.1086/320343},
   DOI={10.1086/320343},
   number={2},
   journal={The Astrophysical Journal},
   publisher={American Astronomical Society},
   author={King, A. R. and Davies, M. B. and Ward, M. J. and Fabbiano, G. and Elvis, M.},
   year={2001},
   month=may, pages={L109–L112} }

@article{Lasota_2023,
   title={Ultraluminous X-ray sources are beamed},
   volume={526},
   ISSN={1365-2966},
   url={http://dx.doi.org/10.1093/mnras/stad2926},
   DOI={10.1093/mnras/stad2926},
   number={2},
   journal={Monthly Notices of the Royal Astronomical Society},
   publisher={Oxford University Press (OUP)},
   author={Lasota, Jean–Pierre and King, Andrew},
   year={2023},
   month=sep, pages={2506–2509}
}

@article{Miller_2003,
   title={X-Ray Spectroscopic Evidence for Intermediate-Mass Black Holes: Cool Accretion Disks in Two Ultraluminous X-Ray Sources},
   volume={585},
   ISSN={1538-4357},
   url={http://dx.doi.org/10.1086/368373},
   DOI={10.1086/368373},
   number={1},
   journal={The Astrophysical Journal},
   publisher={American Astronomical Society},
   author={Miller, J. M. and Fabbiano, G. and Miller, M. C. and Fabian, A. C.},
   year={2003},
   month=mar, pages={L37–L40} 
}

@article{Zhao_2005,
   title={Dark Minihalos with Intermediate Mass Black Holes},
   volume={95},
   ISSN={1079-7114},
   url={http://dx.doi.org/10.1103/PhysRevLett.95.011301},
   DOI={10.1103/physrevlett.95.011301},
   number={1},
   journal={Physical Review Letters},
   publisher={American Physical Society (APS)},
   author={Zhao, HongSheng and Silk, Joseph},
   year={2005},
   month=jun 
}

@misc{goulding2023uncover,
      title={UNCOVER: The growth of the first massive black holes from JWST/NIRSpec -- spectroscopic redshift confirmation of an X-ray luminous AGN at z=10.1}, 
      author={Andy D. Goulding and Jenny E. Greene and David J. Setton and Ivo Labbe and Rachel Bezanson and Tim B. Miller and Hakim Atek and Akos Bogdan and Gabriel Brammer and Iryna Chemerynska and Sam E. Cutler and Pratika Dayal and Yoshinobu Fudamoto and Seiji Fujimoto and Lukas J. Furtak and Vasily Kokorev and Gourav Khullar and Joel Leja and Danilo Marchesini and Priyamvada Natarajan and Erica Nelson and Pascal A. Oesch and Richard Pan and Casey Papovich and Sedona H. Price and Pieter van Dokkum and Bingjie Wang and John R. Weaver and Katherine E. Whitaker and Adi Zitrin},
      year={2023},
      eprint={2308.02750},
      archivePrefix={arXiv},
      primaryClass={astro-ph.GA}
}

@misc{natarajan2023detection,
      title={First Detection of an Over-Massive Black Hole Galaxy UHZ1: Evidence for Heavy Black Hole Seed Formation from Direct Collapse}, 
      author={Priyamvada Natarajan and Fabio Pacucci and Angelo Ricarte and Akos Bogdan and Andy D. Goulding and Nico Cappelluti},
      year={2023},
      eprint={2308.02654},
      archivePrefix={arXiv},
      primaryClass={astro-ph.HE}
}

@misc{bogdan2023evidence,
      title={Evidence for heavy seed origin of early supermassive black holes from a z~10 X-ray quasar}, 
      author={Akos Bogdan and Andy Goulding and Priyamvada Natarajan and Orsolya Kovacs and Grant Tremblay and Urmila Chadayammuri and Marta Volonteri and Ralph Kraft and William Forman and Christine Jones and Eugene Churazov and Irina Zhuravleva},
      year={2023},
      eprint={2305.15458},
      archivePrefix={arXiv},
      primaryClass={astro-ph.GA}
}

@article{Windhorst_2022,
   title={JWST PEARLS. Prime Extragalactic Areas for Reionization and Lensing Science: Project Overview and First Results},
   volume={165},
   ISSN={1538-3881},
   url={http://dx.doi.org/10.3847/1538-3881/aca163},
   DOI={10.3847/1538-3881/aca163},
   number={1},
   journal={The Astronomical Journal},
   publisher={American Astronomical Society},
   author={Windhorst, Rogier A. and Cohen, Seth H. and Jansen, Rolf A. and Summers, Jake and Tompkins, Scott and Conselice, Christopher J. and Driver, Simon P. and Yan, Haojing and Coe, Dan and Frye, Brenda and Grogin, Norman and Koekemoer, Anton and Marshall, Madeline A. and O’Brien, Rosalia and Pirzkal, Nor and Robotham, Aaron and Ryan, Russell E. and Willmer},
   year={2022},
   month=dec, pages={13}
}

@article{Madau_2017,
   title={Radiation Backgrounds at Cosmic Dawn: X-Rays from Compact Binaries},
   volume={840},
   ISSN={1538-4357},
   url={http://dx.doi.org/10.3847/1538-4357/aa6af9},
   DOI={10.3847/1538-4357/aa6af9},
   number={1},
   journal={The Astrophysical Journal},
   publisher={American Astronomical Society},
   author={Madau, Piero and Fragos, Tassos},
   year={2017},
   month=may, pages={39} 
}

@misc{Abbott_2022,
      title={The population of merging compact binaries inferred using gravitational waves through GWTC-3}, 
      author={{LIGO Scientific Collaboration, Virgo Collaboration and KAGRA Collaboration}},
      year={2022},
      eprint={2111.03634},
      archivePrefix={arXiv},
      primaryClass={astro-ph.HE}
}

@article{Tanikawa_2021,
   title={Population III binary black holes: effects of convective overshooting on formation of GW190521},
   volume={505},
   ISSN={1365-2966},
   url={http://dx.doi.org/10.1093/mnras/stab1421},
   DOI={10.1093/mnras/stab1421},
   number={2},
   journal={Monthly Notices of the Royal Astronomical Society},
   publisher={Oxford University Press (OUP)},
   author={Tanikawa, Ataru and Kinugawa, Tomoya and Yoshida, Takashi and Hijikawa, Kotaro and Umeda, Hideyuki},
   year={2021},
   month=may, pages={2170–2176} 
}

@article{Tanikawa_2022,
   title={Merger Rate Density of Binary Black Holes through Isolated Population I, II, III and Extremely Metal-poor Binary Star Evolution},
   volume={926},
   ISSN={1538-4357},
   url={http://dx.doi.org/10.3847/1538-4357/ac4247},
   DOI={10.3847/1538-4357/ac4247},
   number={1},
   journal={The Astrophysical Journal},
   publisher={American Astronomical Society},
   author={Tanikawa, Ataru and Yoshida, Takashi and Kinugawa, Tomoya and Trani, Alessandro A. and Hosokawa, Takashi and Susa, Hajime and Omukai, Kazuyuki},
   year={2022},
   month=feb, pages={83} 
}

@article{Skinner_2020,
   title={Cradles of the first stars: self-shielding, halo masses, and multiplicity},
   volume={492},
   ISSN={1365-2966},
   url={http://dx.doi.org/10.1093/mnras/staa139},
   DOI={10.1093/mnras/staa139},
   number={3},
   journal={Monthly Notices of the Royal Astronomical Society},
   publisher={Oxford University Press (OUP)},
   author={Skinner, Danielle and Wise, John H},
   year={2020},
   month=jan, pages={4386–4397} 
}

@article{Jaacks_2019,
   title={Legacy of star formation in the pre-reionization universe},
   volume={488},
   ISSN={1365-2966},
   url={http://dx.doi.org/10.1093/mnras/stz1529},
   DOI={10.1093/mnras/stz1529},
   number={2},
   journal={Monthly Notices of the Royal Astronomical Society},
   publisher={Oxford University Press (OUP)},
   author={Jaacks, Jason and Finkelstein, Steven L and Bromm, Volker},
   year={2019},
   month=jun, pages={2202–2221} 
}

@article{Hartwig_2022,
   title={Public Release of A-SLOTH: Ancient Stars and Local Observables by Tracing Halos},
   volume={936},
   ISSN={1538-4357},
   url={http://dx.doi.org/10.3847/1538-4357/ac7150},
   DOI={10.3847/1538-4357/ac7150},
   number={1},
   journal={The Astrophysical Journal},
   publisher={American Astronomical Society},
   author={Hartwig, Tilman and Magg, Mattis and Chen, Li-Hsin and Tarumi, Yuta and Bromm, Volker and Glover, Simon C. O. and Ji, Alexander P. and Klessen, Ralf S. and Latif, Muhammad A. and Volonteri, Marta and Yoshida, Naoki},
   year={2022},
   month=aug, pages={45} 
}

@article{Hartwig_2024,
   title={<scp>a-sloth</scp> reveals the nature of the first stars},
   volume={535},
   ISSN={1365-2966},
   url={http://dx.doi.org/10.1093/mnras/stae2318},
   DOI={10.1093/mnras/stae2318},
   number={1},
   journal={Monthly Notices of the Royal Astronomical Society},
   publisher={Oxford University Press (OUP)},
   author={Hartwig, Tilman and Lipatova, Veronika and Glover, Simon C O and Klessen, Ralf S},
   year={2024},
   month=oct, pages={516–530} 
}

@article{Jiao_2023,
   title={Detection of the Keplerian decline in the Milky Way rotation curve},
   volume={678},
   ISSN={1432-0746},
   url={http://dx.doi.org/10.1051/0004-6361/202347513},
   DOI={10.1051/0004-6361/202347513},
   journal={Astronomy \& Astrophysics},
   publisher={EDP Sciences},
   author={Jiao, Yongjun and Hammer, François and Wang, Haifeng and Wang, Jianling and Amram, Philippe and Chemin, Laurent and Yang, Yanbin},
   year={2023},
   month=oct, pages={A208} 
}

@misc{aschersleben2024gamma,
      title={Gamma rays from dark matter spikes in EAGLE simulations}, 
      author={J. Aschersleben and G. Bertone and D. Horns and E. Moulin and R. F. Peletier and M. Vecchi},
      year={2024},
      eprint={2401.14072},
      archivePrefix={arXiv},
      primaryClass={astro-ph.HE}
}

@article{Fragione_2023,
   title={Constraining the Cosmic Merger History of Intermediate-mass Black Holes with Gravitational Wave Detectors},
   volume={944},
   ISSN={1538-4357},
   url={http://dx.doi.org/10.3847/1538-4357/acb34e},
   DOI={10.3847/1538-4357/acb34e},
   number={1},
   journal={The Astrophysical Journal},
   publisher={American Astronomical Society},
   author={Fragione, Giacomo and Loeb, Abraham},
   year={2023},
   month=feb, pages={81} 
}

@article{Fryer_2012,
   title={COMPACT REMNANT MASS FUNCTION: DEPENDENCE ON THE EXPLOSION MECHANISM AND METALLICITY},
   volume={749},
   ISSN={1538-4357},
   url={http://dx.doi.org/10.1088/0004-637X/749/1/91},
   DOI={10.1088/0004-637x/749/1/91},
   number={1},
   journal={The Astrophysical Journal},
   publisher={American Astronomical Society},
   author={Fryer, Chris L. and Belczynski, Krzysztof and Wiktorowicz, Grzegorz and Dominik, Michal and Kalogera, Vicky and Holz, Daniel E.},
   year={2012},
   month=mar, pages={91} 
}

@article{Belczynski_2016,
   title={The effect of pair-instability mass loss on black-hole mergers},
   volume={594},
   ISSN={1432-0746},
   url={http://dx.doi.org/10.1051/0004-6361/201628980},
   DOI={10.1051/0004-6361/201628980},
   journal={Astronomy \& Astrophysics},
   publisher={EDP Sciences},
   author={Belczynski, K. and Heger, A. and Gladysz, W. and Ruiter, A. J. and Woosley, S. and Wiktorowicz, G. and Chen, H.-Y. and Bulik, T. and O’Shaughnessy, R. and Holz, D. E. and Fryer, C. L. and Berti, E.},
   year={2016},
   month=oct, pages={A97} 
}

@article{Woosley_2017,
   title={Pulsational Pair-instability Supernovae},
   volume={836},
   ISSN={1538-4357},
   url={http://dx.doi.org/10.3847/1538-4357/836/2/244},
   DOI={10.3847/1538-4357/836/2/244},
   number={2},
   journal={The Astrophysical Journal},
   publisher={American Astronomical Society},
   author={Woosley, S. E.},
   year={2017},
   month=feb, pages={244}
}

@article{Renzo_2020,
   title={Predictions for the hydrogen-free ejecta of pulsational pair-instability supernovae},
   volume={640},
   ISSN={1432-0746},
   url={http://dx.doi.org/10.1051/0004-6361/202037710},
   DOI={10.1051/0004-6361/202037710},
   journal={Astronomy \& Astrophysics},
   publisher={EDP Sciences},
   author={Renzo, M. and Farmer, R. and Justham, S. and Götberg, Y. and de Mink, S. E. and Zapartas, E. and Marchant, P. and Smith, N.},
   year={2020},
   month=aug, pages={A56} 
}

@article{Fryer_2001,
   title={Pair‐Instability Supernovae, Gravity Waves, and Gamma‐Ray Transients},
   volume={550},
   ISSN={1538-4357},
   url={http://dx.doi.org/10.1086/319719},
   DOI={10.1086/319719},
   number={1},
   journal={The Astrophysical Journal},
   publisher={American Astronomical Society},
   author={Fryer, C. L. and Woosley, S. E. and Heger, A.},
   year={2001},
   month=mar, pages={372–382} 
}

@article{Hill_2018,
   title={The Spectrum of the Universe},
   volume={72},
   ISSN={1943-3530},
   url={http://dx.doi.org/10.1177/0003702818767133},
   DOI={10.1177/0003702818767133},
   number={5},
   journal={Applied Spectroscopy},
   publisher={SAGE Publications},
   author={Hill, Ryley and Masui, Kiyoshi W. and Scott, Douglas},
   year={2018},
   month=apr, pages={663–688} 
}

@article{H_berle_2024,
   title={Fast-moving stars around an intermediate-mass black hole in ω Centauri},
   volume={631},
   ISSN={1476-4687},
   url={http://dx.doi.org/10.1038/s41586-024-07511-z},
   DOI={10.1038/s41586-024-07511-z},
   number={8020},
   journal={Nature},
   publisher={Springer Science and Business Media LLC},
   author={Häberle, Maximilian and Neumayer, Nadine and Seth, Anil and Bellini, Andrea and Libralato, Mattia and Baumgardt, Holger and Whitaker, Matthew and Dumont, Antoine and Alfaro-Cuello, Mayte and Anderson, Jay and Clontz, Callie and Kacharov, Nikolay and Kamann, Sebastian and Feldmeier-Krause, Anja and Milone, Antonino and Nitschai, Maria Selina and Pechetti, Renuka and van de Ven, Glenn},
   year={2024},
   month=jul, pages={285–288} }

@ARTICLE{2024arXiv240716373P,
       author = {{Postnov}, Konstantin and {Chekh}, Ilya},
        title = "{Primordial Intermediate-mass Binary Black Holes as Targets for Space Laser Interferometers}",
      journal = {arXiv e-prints},
         year = 2024,
        month = jul,
          eid = {arXiv:2407.16373},
        pages = {arXiv:2407.16373},
          doi = {10.48550/arXiv.2407.16373},
archivePrefix = {arXiv},
       eprint = {2407.16373},
 primaryClass = {astro-ph.CO},
       adsurl = {https://ui.adsabs.harvard.edu/abs/2024arXiv240716373P}
}

@ARTICLE{2016JCAP...11..036B,
       author = {{Blinnikov}, S. and {Dolgov}, A. and {Porayko}, N.~K. and {Postnov}, K.},
        title = "{Solving puzzles of GW150914 by primordial black holes}",
      journal = {\jcap},
         year = 2016,
        month = nov,
       volume = {2016},
       number = {11},
          eid = {036},
        pages = {036},
          doi = {10.1088/1475-7516/2016/11/036},
archivePrefix = {arXiv},
       eprint = {1611.00541},
 primaryClass = {astro-ph.HE},
       adsurl = {https://ui.adsabs.harvard.edu/abs/2016JCAP...11..036B}
}

@misc{klessen2023,
      title={The first stars: formation, properties, and impact}, 
      author={Ralf S. Klessen and Simon C. O. Glover},
      year={2023},
      eprint={2303.12500},
      archivePrefix={arXiv},
      primaryClass={astro-ph.CO},
      url={https://arxiv.org/abs/2303.12500}, 
}

@article{Cai_2025,
   title={A Metal-free Galaxy at z = 3.19? Evidence of Late Population III Star Formation at Cosmic Noon},
   volume={993},
   ISSN={2041-8213},
   url={http://dx.doi.org/10.3847/2041-8213/ae1608},
   DOI={10.3847/2041-8213/ae1608},
   number={2},
   journal={The Astrophysical Journal Letters},
   publisher={American Astronomical Society},
   author={Cai, Sijia and Li, Mingyu and Cai, Zheng and Wu, Yunjing and Yu, Fujiang and Dickinson, Mark and Sun, Fengwu and Fan, Xiaohui and Ben Wang and Cullen, Fergus and Bian, Fuyan and Lin, Xiaojing and Zou, Jiaqi},
   year={2025},
   month=nov, pages={L52} }

@article{Sun_2021,
   title={Revealing the formation histories of the first stars with the cosmic near-infrared background},
   volume={508},
   ISSN={1365-2966},
   url={http://dx.doi.org/10.1093/mnras/stab2697},
   DOI={10.1093/mnras/stab2697},
   number={2},
   journal={Monthly Notices of the Royal Astronomical Society},
   publisher={Oxford University Press (OUP)},
   author={Sun, Guochao and Mirocha, Jordan and Mebane, Richard H and Furlanetto, Steven R},
   year={2021},
   month=sep, pages={1954–1972} }

\end{document}